\documentclass[10pt,confenrence]{IEEEtran}
\IEEEoverridecommandlockouts

\usepackage{cite}
\usepackage{amsmath,amssymb,amsfonts}
\usepackage{algorithmic}
\usepackage{graphicx}
\usepackage{textcomp}
\usepackage{xcolor}
\usepackage{url}
\usepackage{enumitem}
\usepackage{multirow}
\usepackage{booktabs}

\def\BibTeX{{\rm B\kern-.05em{\sc i\kern-.025em b}\kern-.08em
    T\kern-.1667em\lower.7ex\hbox{E}\kern-.125emX}}
\begin{document}

\title{Self-Distillation Prototypes Network: Learning Robust Speaker Representations without Supervision}

\author{\IEEEauthorblockN{
Yafeng Chen, Siqi Zheng, Hui Wang, Luyao Cheng, Qian Chen, Chong Deng, Shiliang Zhang, Wen Wang}
\\
\IEEEauthorblockA{\textit{Speech Lab, Alibaba Group}}
}

\maketitle

\begin{abstract}
Training speaker-discriminative and robust speaker verification systems without explicit speaker labels remains a persistent challenge. In this paper, we propose a novel self-supervised speaker verification approach, \textbf{Self-Distillation Prototypes Network (SDPN)}, which effectively facilitates self-supervised speaker representation learning. SDPN assigns the representation of the augmented views of an utterance to the same prototypes as the representation of the original view, thereby enabling effective knowledge transfer between the augmented and original views. Due to lack of negative pairs in the SDPN training process, the network tends to align positive pairs quite closely in the embedding space, a phenomenon known as \textit{model collapse}. To mitigate this problem, we introduce a \textit{diversity regularization term} to embeddings in SDPN. Comprehensive experiments on the VoxCeleb datasets demonstrate the superiority of SDPN among self-supervised speaker verification approaches. SDPN sets a new state-of-the-art on the VoxCeleb1 speaker verification evaluation benchmark, achieving Equal Error Rate \textbf{1.80\%}, \textbf{1.99\%}, and \textbf{3.62\%} for trial VoxCeleb1-O, VoxCeleb1-E and VoxCeleb1-H respectively\footnote{Code is publicly available at \url{https://github.com/modelscope/3D-Speaker}}, without using any speaker labels in training.
Ablation studies show that both proposed learnable prototypes in self-distillation network and diversity regularization contribute to the verification performance.
\end{abstract}


\section{Introduction}
\label{sec:intro}
Speaker verification (SV) systems based on deep learning have achieved remarkable progress in recent years. The availability of large-scale labeled datasets is critical for the performance of the deep learning systems. However, collecting large amounts of labeled real-world SV data is laborious and expensive. Therefore, it is of great interest to explore approaches reducing the reliance on labeled data, such as self-supervised learning (SSL) approaches.

SSL methods learn representations of data without supervision (e.g., class labels). They can be roughly categorized into contrastive~\cite{ DBLP:conf/icml/ChenK0H20, DBLP:journals/corr/abs-1807-03748, DBLP:conf/cvpr/He0WXG20, DBLP:conf/icassp/ZhangZW21, DBLP:conf/icassp/ZhangJCLHS22,DBLP:conf/nips/0004LW023,DBLP:journals/taslp/TuMC24} and non-contrastive approaches~\cite{DBLP:conf/icassp/SangLLAW22, DBLP:conf/nips/GrillSATRBDPGAP20, DBLP:conf/iccv/CaronTMJMBJ21, DBLP:conf/interspeech/HeoJKKLKC23, chen2023pushing, DBLP:conf/slt/ChenQHQZ22, DBLP:journals/corr/abs-2304-05754, DBLP:conf/icassp/TaoLDHL22,zhou2024self,jinself,citation-0}, depending on whether the training process uses negative samples or not. Contrastive methods require large batch sizes or special techniques such as memory banks to attain high performance. More importantly, the quality of training is highly dependent on the correctness of negative pairs. 
For SSL based SV, since speaker labels are unavailable, a common practice is to construct positive pairs from the same utterance and negative pairs from different utterances; hence, within a batch, negative pairs may come from the same speaker and would be incorrect.

In contrast, non-contrastive SSL methods do not have this issue since negative samples are not required in the training process, yet non-contrastive SSL methods have shown comparable or better performance compared to contrastive counterparts~\cite{DBLP:conf/icassp/SangLLAW22, DBLP:conf/nips/GrillSATRBDPGAP20, DBLP:conf/iccv/CaronTMJMBJ21, DBLP:conf/interspeech/HeoJKKLKC23, chen2023pushing, DBLP:conf/slt/ChenQHQZ22, DBLP:journals/corr/abs-2304-05754, DBLP:conf/icassp/TaoLDHL22,zhou2024self,jinself,citation-0}. Non-contrastive SSL methods can be broadly categorized into single-stage training~\cite{DBLP:conf/icassp/SangLLAW22, DBLP:conf/nips/GrillSATRBDPGAP20, DBLP:conf/iccv/CaronTMJMBJ21, DBLP:conf/interspeech/HeoJKKLKC23, chen2023pushing, DBLP:conf/slt/ChenQHQZ22,jinself,citation-0} and multi-stage training methods~\cite{DBLP:journals/corr/abs-2304-05754, DBLP:conf/icassp/TaoLDHL22,zhou2024self}. Differences between the single-stage and multi-stage training methods are detailed in~\cite{chen2023pushing}. 

\begin{figure*}[thb]
  \centering
  \includegraphics[scale=0.44]{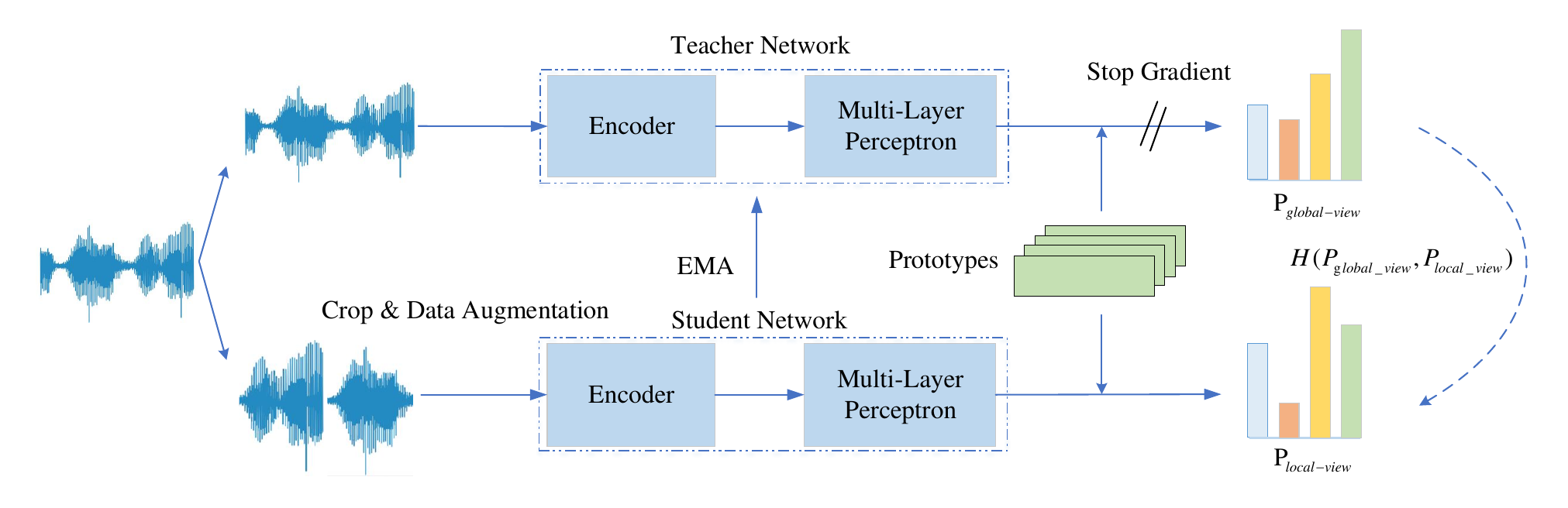}
    \vspace{-2mm}
    \caption{Overview of our proposed \textbf{Self-Distillation Prototypes Network (SDPN)}. SDPN comprises a teacher network and a student network with identical architecture but different parameters. The teacher/student network consists of three main modules: the \textbf{encoder} for extracting speaker embeddings, a \textbf{multi-layer perceptron} for feature transformation, and learnable \textbf{prototypes} for computing soft distributions between global and local views. Global views and local views refer to the long and short segments randomly segmented from the same utterance, respectively. \textbf{EMA} denotes Exponential Moving Average.}
    \label{fig:overview}
\end{figure*}

Note that the multi-stage training methods require an estimation of the rough number of speakers in the entire dataset and commonly use a clustering algorithm to generate pseudo speaker labels
for each utterance, which contradicts the original purpose of SSL. 
Consequently, in this work, we adopt the line of \textbf{single-stage non-contrastive approaches} for self-supervised SV. Among prior works in this direction, Sang et al.~\cite{DBLP:conf/icassp/SangLLAW22} propose a regularization structure inspired by BYOL~\cite{DBLP:conf/nips/GrillSATRBDPGAP20}. In BYOL, an online network predicts a target network representation of the same utterance under different augmented views. Caron et al.~\cite{DBLP:conf/iccv/CaronTMJMBJ21} propose a self-distillation framework DINO that comprises a teacher and a student network. The outputs of the teacher network are used as ground truth to optimize the student network. Heo et al. \cite{DBLP:conf/interspeech/HeoJKKLKC23} further introduce a curriculum learning strategy to DINO to guide model training.
Due to lack of negative pairs, these non-contrastive SSL methods are inclined to map positive pairs to the same or similar positions in the embedding space, resulting in degenerate solutions, a phenomenon known as \textbf{model collapse}.  
To alleviate model collapse, Chen et al.~\cite{chen2023pushing} propose two regularization terms in DINO. \cite{DBLP:conf/slt/ChenQHQZ22} enhance SV performance by applying augmentation strategies to DINO.  

Although these non-contrastive methods~\cite{DBLP:conf/icassp/SangLLAW22, DBLP:conf/slt/ChenQHQZ22, DBLP:journals/jstsp/ZhangY22} demonstrate promising speaker verification performance of applying DINO and BYOL based approaches for self-supervised speaker representation learning, the key limitation of these approaches lies in their emphasis on minimizing the distance between representations of the same utterance\footnote{In self-supervised SV, since there are no speaker labels, the short segments from the same utterance are considered belonging to the same speaker, and different utterances are considered belonging to different speakers. However, different utterances may belong to the same speaker.}
, while overlooking the relationship between representations of utterances from different speakers, which substantially limits the model capability.

To tackle this limitation, we propose a novel non-contrastive \textbf{Self-Distillation Prototypes Network (SDPN)} for self-supervised speaker verification. In order to alleviate model collapse within the SDPN framework, we also introduce a regularization term to embeddings. Our contributions can be summarized as follows:

\begin{itemize}[leftmargin=*]
    \item We propose a novel Self-Distillation Prototypes Network (\textbf{SDPN}) for self-supervised speaker verification. Our first key innovation is \textbf{introducing learnable prototypes in self-distillation framework} to utilize unlabeled data to learn robust speaker-discriminative embeddings. The role of the learnable prototypes is similar to that of a classifier in fully supervised learning.
    SDPN explicitly minimizes the distance between different augmented views of the same utterance and concurrently, implicitly infuses the membership information by assigning representations of different utterances to their corresponding basis vectors in prototypes. In this way, SDPN enhances the compactness of embeddings for the same utterance while separating embeddings from different speakers, effectively addressing the limitation of prior non-contrastive SSL approaches.

    \item Our second key innovation is proposing \textbf{a diversity regularization term} and integrating it within SDPN. This component computes the pairwise similarity among embeddings and actively segregates the nearest embeddings to promote the diversity of speaker embeddings within a batch. Incorporating diversity regularization further enhances the robustness of speaker embeddings.

    \item Experiments on the VoxCeleb1 speaker verification evaluation benchmark demonstrate the superiority of our SDPN for self-supervised SV. SDPN substantially narrows the performance gap between SSL and supervised methods for SV and \textbf{sets a new state of the art (SOTA)} on VoxCeleb1, achieving equal error rate  \textbf{1.80\%}, \textbf{1.99\%}, and \textbf{3.62\%} for trial VoxCeleb1-O, VoxCeleb1-E, and VoxCeleb1-H, without using any speaker labels in training. Ablation studies show that both proposed learnable prototypes in self-distillation network and diversity regularization contribute to the speaker verification performance.
\end{itemize}

\begin{table*}[thb]
    \caption{Results on VoxCeleb1-O, VoxCeleb1-E, and VoxCeleb1-H datasets. DINO* refers to our replication of the baseline DINO framework. SDPN w/o PRO denotes removing prototypes from our proposed SDPN. SDPN w/o PRO-2 denotes removing prototypes but increasing the number of parameters in the MLP layers, resulting in the same model size as SDPN. SDPN w/ DR denotes adding the diversity regularization to SDPN. The best results for each test set are in bold.}
    \vspace{-2mm}
    \label{tab:main_results}
    \centering
    \begin{tabular}{c c c c c c c c}
    \toprule
    \multirow{2}{*}{\textbf{Architecture}} & \multirow{2}{*}{\textbf{Params (M)}} & \multicolumn{2}{c}{\textbf{VoxCeleb1-O}} & \multicolumn{2}{c}{\textbf{VoxCeleb1-E}} & \multicolumn{2}{c}{\textbf{VoxCeleb1-H}} \\
    \cmidrule(lr){3-4} \cmidrule(lr){5-6} \cmidrule(lr){7-8}
    & & \textbf{EER (\%)}$\downarrow$ & \textbf{minDCF}$\downarrow$ & \textbf{EER (\%)}$\downarrow$ & \textbf{minDCF}$\downarrow$ & \textbf{EER (\%)}$\downarrow$ & \textbf{minDCF}$\downarrow$ \\
    \midrule
    DINO* & 90.68 & 2.65 & 0.202 & 2.74 & 0.188 & 5.02 & 0.304 \\
    \midrule
    SDPN & 57.24 & 2.13 & 0.170 & 2.12 & 0.150 & 4.15 & 0.263 \\
    SDPN w/o PRO & 56.98 & 3.76 & 0.278 & 3.93 & 0.271 & 7.28 & 0.436 \\
    SDPN w/o PRO-2 & 57.29 & 3.75 & 0.274 & 3.90 & 0.270 & 7.22 & 0.433 \\
    \textbf{SDPN w/ DR} & 57.24 & \textbf{1.80} & \textbf{0.139} & \textbf{1.99} & \textbf{0.131} & \textbf{3.62} & \textbf{0.219} \\
    \bottomrule
    \end{tabular}
\end{table*}

\section{Proposed Method}
\label{sec:method}
Fig.~\ref{fig:overview} illustrates the architecture of the proposed Self-Distillation Prototypes Network.
The upper branch in the figure presents the teacher network, while the lower branch shows the student network. Both networks share the same architecture with different parameters. 
The teacher/student network comprises an encoder \emph{f} for extracting speaker embeddings and a multi-layer perceptron (MLP) \emph{h} for conducting a non-linear transformation of these embeddings. We adopt ECAPA-TDNN~\cite{DBLP:conf/interspeech/DesplanquesTD20} as the encoder. The MLP \emph{h} comprises three fully connected layers with hidden dimensions of 2048-2048-256, followed by L2 normalization. The output of the first two layers passes through both batch normalization~\cite{DBLP:conf/icml/IoffeS15} and GELU activation functions~\cite{DBLP:conf/nips/KlambauerUMH17}. The learnable prototypes $\textbf{C}$ (each prototype is an $d$-dimensional vector) are shared between the teacher network and the student network, and are used to compute the soft distributions between global and local views. Note that global views and local views refer to the long and short segments randomly segmented from the same utterance, respectively. 
Global views are fed into the teacher network and local views are fed into the student network. The global information learned by the teacher network guides training of the student network, thereby enforcing the local-global correspondences.

Specifically, we adopt a multi-crop strategy to sample one global view $\textbf{X}_g = \{\textbf{x}_{g}\}$ and four local views $\textbf{X}_l = \{\textbf{x}_{l_1}, \textbf{x}_{l_2}, \textbf{x}_{l_3}, \textbf{x}_{l_4}\}$ from an utterance. 
Data augmentation have been proven to be crucial for both supervised and self-supervised representation learning. In order to adequately capture the utterance-dependent variability in speaker embeddings, we explore two types of augmentation strategies, WavAugment~\cite{DBLP:conf/icassp/SnyderGSPK18} and SpecAugment~\cite{DBLP:conf/interspeech/ParkCZCZCL19} on the local views in SDPN, while the global view is left unchanged to ensure the integrity of the utterance. 
Then, $\textbf{X}_g$ is first encoded by the teacher's encoder $f^{tea}_{\vartheta}$ and the resulting representations are taken as speaker embeddings. Next, speaker embeddings are mapped through the teacher's MLP $h^{tea}_{\vartheta}$. At the same time, the four local views $\textbf{X}_l$ are encoded by the student's encoder $f^{stu}_{\theta}$ and then mapped through the student's MLP $h^{stu}_{\theta}$. The parameters $\vartheta$ of the teacher encoder and MLP are updated via Exponential Moving Average (\textbf{EMA}) of the parameters $\theta$ of the student encoder and MLP. To train the encoders, we compute the probability distributions between all prototypes $\textbf{C}$ and each pair of local view and global view.
The cross-entropy (CE) loss is calculated to minimize the probability distribution as follows:
\begin{equation}
    \mathcal{L}_{CE} = \sum_{\textbf{x} \in \textbf{X}_{g}}\sum_{\substack{\textbf{x}^{\prime} \in \mathbf{X}_{l}}} H(P^{tea}(\textbf{x}) \mid P^{stu}(\textbf{x}^{\prime}))
\end{equation}
\begin{equation}
    P^{tea}(\textbf{x})=Sknorm\left(h_{\vartheta}^{tea}(f_{\vartheta}^{tea}(\textbf{x})) \cdot \textbf{C} / \tau_t\right)
\end{equation}
\begin{equation}
    P^{stu}(\textbf{x}^{\prime})=Softmax \left(h_{\theta}^{stu}(f_{\theta}^{stu}(\textbf{x}^{\prime})) \cdot \textbf{C} / \tau_s\right)
\end{equation}
where $H(a | b)=-a*\log b$ is cross-entropy. $P^{tea}$ and $P^{stu}$ denote the output probability distributions of the teacher network and the student network, where $\tau_t$ and $\tau_s$ are temperature parameters that control the sharpness of the teacher's and student's output distributions (lower temperature for sharper output). $Sknorm$ denotes Sinkhorn-Knopp (SK) batch normalization~\cite{DBLP:conf/nips/Cuturi13}, which helps stabilize the teacher network. Softmax is applied to the student's output.

We design a diversity regularization term to enhance the diversity of speaker embeddings within a batch. This regularizer forces the embeddings of utterances to be different, thereby preventing trivial solutions and alleviating model collapse. First, we compute the pairwise similarity of speaker embeddings($\textbf{x}_i$ and $\textbf{x}_j$) within a batch. Next, all the closest embeddings are separated in order to enhance their dissimilarity. The diversity regularization loss is calculated as follows:
\begin{equation}
\label{eq:diversity_regularization}
    \mathcal{L}_{DR} = -\frac{1}{n} \sum_{i=1}^n (\sum_{j=1}^n \log(\min_{j \neq i} || \mathbf{x}_i - \mathbf{x}_j ||))
\end{equation}
where $n$ is the batch size. Diversity regularization implicitly considers relationship between different speakers
and further complements SDPN. The overall training objective is to minimize a combination of the CE loss and diversity regularization loss, weighted by the hyperparameter $\mu$.
\begin{equation}
\label{eq:overall_loss}
  \mathcal{L} = \mathcal{L}_{CE} + \mu \mathcal{L}_{DR}
\end{equation}
%

\section{Experiments}
\label{sec:experiments}
\subsection{Experimental Setup}
\noindent \textbf{Datasets and Evaluation Metrics}
We evaluate the efficacy of the proposed approach on VoxCeleb datasets. Specifically, we adopt the development portions of VoxCeleb2~\cite{DBLP:conf/interspeech/ChungNZ18} for training, which comprises 1,092,009 utterances across 5,994 speakers.
No speaker labels are used in training in any experiment. 
We report the results on
three trials on VoxCeleb1~\cite{DBLP:conf/interspeech/NagraniCZ17}, 
in terms of two metrics, namely, equal error rate (EER) and the minimum of the normalized detection cost function (minDCF) with the settings of $P_{target}$ = 0.05 and $C_{fa} = C_{miss}$ = 1.

\noindent \textbf{Input Features}
For each utterance, we use the multi-crop strategy (Section~\ref{sec:method}) for SDPN training in which one 4s segment and four 2s segments are sampled from an utterance as its global view and local views, respectively. The acoustic features used in the experiments are 80-dimensional Filter Bank (FBank) with 25ms windows and 10ms shift. 

\noindent \textbf{Data Augmentation} We explore WavAugment and SpecAugment in SDPN. For WavAugment, the MUSAN corpus~\cite{DBLP:journals/corr/SnyderCP15} with SNR 0-15 for additive noise and Room Impulse Response (RIR) \cite{DBLP:conf/icassp/KoPPSK17} for reverberation are randomly applied to each local view. For SpecAugment, one time mask and one frequency mask are randomly applied to the FBank features of the local views. The time masking length is 0-10 frames and the frequency masking length is 0-6 dimensions. 

\noindent \textbf{Implementation Details} We use ECAPA-TDNN with attentive statistical pooling as the encoder. The weight decay is fixed as 5e-5. The learning rate scheduling starts with 10 warm-up epochs with a linear increase from 0 to 0.4, followed by a cosine decay with a final learning rate of 1e-5. We train the model with 150 epochs using the stochastic gradient descent optimizer~\cite{DBLP:journals/ijon/Amari93} with a momentum of 0.9, on 4 NVIDIA V100 GPUs. The temperature $\tau_t$ and $\tau_s$ are set to 0.04 and 0.1, respectively. $\mu$ in Eq.\ref{eq:overall_loss} is set to 0.1. Each prototype has a dimension $d$ of 256, and the number of prototypes is 1024. The dimension of speaker embeddings is 512.

\vspace{-4mm}
\subsection{Results and Analysis}
DINO is currently the most prevalent SSL framework for SV, with the majority of competitive models, including the current SSL SOTA C3-DINO~\cite{DBLP:journals/jstsp/ZhangY22}, based on DINO. Hence, we compare the results of our SDPN and our replicated baseline DINO\footnote{Our reproduced DINO achieved a better 2.65\% EER than 3.30\% EER reported for DINO in the C3-DINO paper~\cite{DBLP:journals/jstsp/ZhangY22} on VoxCeleb1-O.} 
on the three VoxCeleb1-\{O,E,H\} test sets, as shown in Table~\ref{tab:main_results}.  Comparing row 1 and 2 shows that \textit{SDPN outperforms DINO substantially and consistently across all test sets}, achieving EERs of \textbf{2.13\%}, \textbf{2.12\%}, and \textbf{4.15\%}, with \textbf{much smaller model size (57.24M)}, which is only \textbf{63\%} of the size of DINO (90.68M). These results demonstrate the superiority of SDPN in self-supervised SV. 

\begin{figure}[thb]
  \centering
  \includegraphics[scale=0.2]{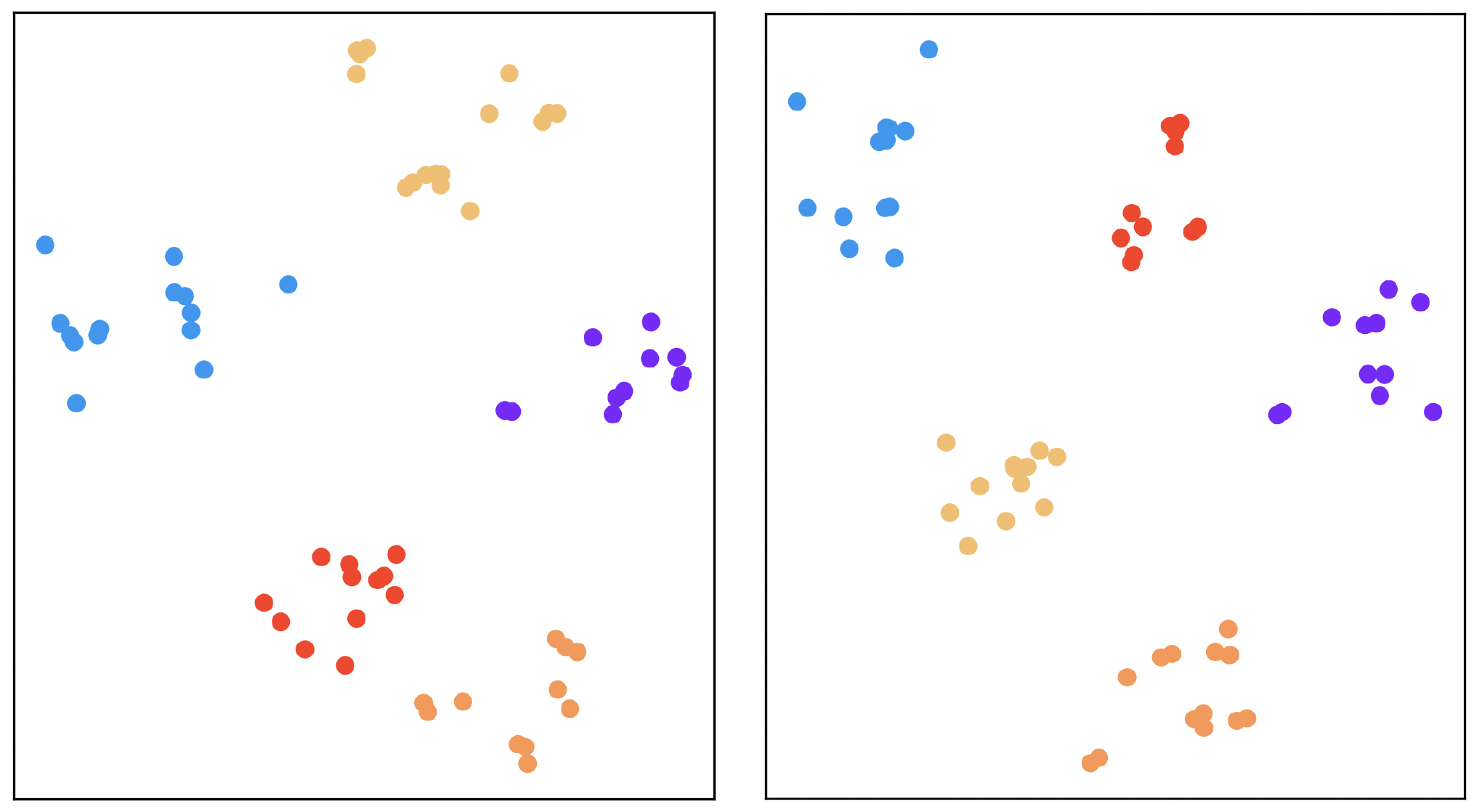}
  \vspace{-2mm}
  \caption{t-SNE visualization of extracted embeddings for five speakers, each denoted by a distinct color. The left figure displays speaker embeddings from baseline DINO, while the right figure shows those from our SDPN with diversity regularization. The embeddings from SDPN with diversity regularization clearly exhibit enhanced separation over those from DINO, suggesting better discriminability.}
  \label{fig:visualization}
\end{figure}

\begin{table}[thb]
    \caption{Comparison between the results of our SDPN w/ DR and results cited from recent SSL models on the VoxCeleb-O.}
    \vspace{-2mm}
    \label{tab:compare_ssl}
    \centering
    \setlength{\tabcolsep}{3pt}
    \begin{tabular}{lp{3.0cm}c}
    \toprule
    \textbf{Model} & \textbf{Embedding Extractor} & \textbf{EER(\%)}$\downarrow$ \\
    \midrule
    SSReg~\cite{DBLP:conf/icassp/SangLLAW22} & Fast ResNet34 & 6.99 \\
    Mixup-Aug~\cite{DBLP:conf/icassp/ZhangJCLHS22} & Fast ResNet34 & 5.84 \\
    DINO + CL \cite{DBLP:conf/interspeech/HeoJKKLKC23} & ECAPA-TDNN & 4.47 \\
    DINO~\cite{DBLP:journals/jstsp/ZhangY22} & ECAPA-TDNN & 3.30 \\
    RDINO~\cite{chen2023pushing} & ECAPA-TDNN & 3.24 \\
    {MeMo-CTES}~\cite{jinself} & ECAPA-TDNN & 3.10 \\
    {RDINO + W-GVKT}~\cite{citation-0} & ECAPA-TDNN & 2.89 \\
    DINO-Aug~\cite{DBLP:conf/slt/ChenQHQZ22} & ECAPA-TDNN & 2.51 \\
    C3-DINO~\cite{DBLP:journals/jstsp/ZhangY22} & ECAPA-TDNN & 2.50 \\
    \midrule
    \textbf{SDPN w/ DR} & \textbf{ECAPA-TDNN} & \textbf{1.80} \\
    \bottomrule
    \end{tabular}
\end{table}

Comparing rows 2 and 3 demonstrates that removing the learnable prototypes results in a notable degradation in SV performance. We hypothesize that the learnable prototypes implicitly capture class membership for different classes without relying on labeled data. Since the prototypes have 0.26M parameters, to understand how much the performance degradation from row 3 over row 2 is due to the smaller model size, we evaluate a counterpart SDPN w/o PRO-2, which compensates the model size reduction from removing prototypes by increasing the number of parameters in the MLP layers by 0.31M, resulting in about the same model size as SDPN. We find that SDPN w/o PRO-2 obtains EER 3.75\% on VoxCeleb1-O, on par with 3.76\% EER from SDPN w/o PRO, yet notably worse than 2.13\% EER from SDPN. Similar observations are made on VoxCeleb1-E and VoxCeleb1-H. These results clearly demonstrate that the \textbf{learnable prototypes contribute substantially to the SV performance}. 

Comparing row 2 and 5 of Table~\ref{tab:main_results} shows that \textbf{adding diversity regularization (SDPN w/ DR)} achieves EERs \textbf{1.80\%}, \textbf{1.99\%}, and \textbf{3.62\%}, with relative gains by \textbf{15.49\%}, \textbf{6.13\%}, and \textbf{12.77\%} on the three test sets, continuously improving speaker verification accuracy.
Diversity regularization also consistently improves minDCF. We employ the t-distributed Stochastic Neighbor Embedding (t-SNE)~\cite{van2008visualizing} to visually compare the disentanglement performance of speaker embeddings derived from DINO and SDPN with diversity regularization, as illustrated in Fig.~\ref{fig:visualization}. It is clear that the embeddings extracted via SDPN w/ DR exhibit superior clustering capabilities compared to those from DINO, suggesting that SDPN makes speaker embeddings more discriminative.

\begin{table}[thb]
    \caption{Effect of diversity regularization weight $\mu$ (Eq.\ref{eq:overall_loss}) on SDPN.}
    \vspace{-2mm}
    \label{tab:weight_effect}
    \centering
    \setlength{\tabcolsep}{0.5pt}
    \fontsize{8.8}{11}\selectfont
    \begin{tabular}{c c c c c c c}
    \toprule
    \multirow{2}{*}{\textbf{Weight}} & \multicolumn{2}{c}{\textbf{VoxCeleb1-O}} & \multicolumn{2}{c}{\textbf{VoxCeleb1-E}} & \multicolumn{2}{c}{\textbf{VoxCeleb1-H}} \\
    \cmidrule(lr){2-3} \cmidrule(lr){4-5} \cmidrule(lr){6-7}
    $\mu$ & \textbf{EER(\%)} & \textbf{minDCF} & \textbf{EER(\%)} & \textbf{minDCF} & \textbf{EER(\%)} & \textbf{minDCF} \\
    \midrule
    0 & 2.13 & 0.170 & 2.12 & 0.150 & 4.15 & 0.263 \\
    0.05 & 1.94 & \textbf{0.134} & 2.02 & 0.132 & 3.67 & 0.227 \\
    0.1 & \textbf{1.80} & 0.139 & \textbf{1.99} & \textbf{0.131} & \textbf{3.62} & \textbf{0.219} \\
    0.2 & 1.95 & 0.157 & 2.08 & 0.136 & 3.74 & 0.223 \\
    \bottomrule
    \end{tabular}
\end{table}

\begin{table}[th]
    \caption{Effect of data augmentation on SDPN with diversity regularization. Aug. denotes Augment. Wav+Spec. denotes the combination of WavAugment and SpecAugment.}
    \vspace{-2mm}
    \label{tab:data_augmentation}
    \centering
    \setlength\tabcolsep{0.5pt}
    \fontsize{8.8}{11}\selectfont
    \begin{tabular}{c c c c c c c}
    \toprule
    \multirow{2}{*}{} & \multicolumn{2}{c}{\textbf{VoxCeleb1-O}} & \multicolumn{2}{c}{\textbf{VoxCeleb1-E}} & \multicolumn{2}{c}{\textbf{VoxCeleb1-H}} \\
    \cmidrule(lr){2-3} \cmidrule(lr){4-5} \cmidrule(lr){6-7}
    & \textbf{EER(\%)} & \textbf{minDCF} & \textbf{EER(\%)} & \textbf{minDCF} & \textbf{EER(\%)} & \textbf{minDCF} \\
    \midrule
    No Aug. & 3.62 & 0.268 & 4.34 & 0.266 & 6.99 & 0.407 \\
    WavAug. & 1.88 & 0.153 & 2.06 & 0.139 & 3.72 & 0.228 \\
    SpecAug. & 4.27 & 0.317 & 5.03 & 0.309 & 7.69 & 0.447 \\
    Wav+Spec. & \textbf{1.80} & \textbf{0.139} & \textbf{1.99} & \textbf{0.131} & \textbf{3.62} & \textbf{0.219} \\
    \bottomrule
    \end{tabular}
\end{table}

We compare SDPN to recently published \textbf{non-contrastive} SSL approaches, including \cite{ DBLP:conf/icassp/ZhangJCLHS22, DBLP:conf/icassp/SangLLAW22, DBLP:conf/interspeech/HeoJKKLKC23, chen2023pushing, DBLP:conf/slt/ChenQHQZ22,jinself,citation-0}, and the \textbf{SSL SOTA C3-DINO}~\cite{DBLP:journals/jstsp/ZhangY22} which \textit{integrates contrastive and non-contrastive methods}, as shown in Table~\ref{tab:compare_ssl}. On VoxCeleb1-O test set, our non-contrastive SDPN with diversity regularization achieves \textbf{1.80\%} EER using the same cosine distance scoring method as C3-DINO, outperforming C3-DINO (2.50\% EER) by \textbf{28.0\%} relative. 

We analyze the impact of the weight of diversity regularization on SV performance, as shown in Table~\ref{tab:weight_effect}. We find that applying diversity regularization outperforms DINO with even a small weight $\mu = 0.05$. SDPN w/ DR achieves \textbf{1.80\%}, \textbf{1.99\%}, and \textbf{3.62\%} EER with $\mu = 0.1$ on VoxCeleb1-O, VoxCeleb1-E, and VoxCeleb1-H.

We also study the effect of different data augmentation strategies on the training data, as shown in Table~\ref{tab:data_augmentation}. We find that when no augmentation is applied, it is difficult for the entire network to converge due to the inherent property of non-contrastive frameworks. WavAugment notably improves SDPN performance. 
Combining WavAugment and SpecAugment outperforms the use of WavAugment alone, indicating that these augmentation strategies complement each other.

\section{Conclusion}
We propose a novel self-distillation prototypes network with diversity regularization for self-supervised learning for speaker verification, utilizing unlabeled data to learn robust speaker-discriminative embeddings. 
SDPN enhances compactness of embeddings from the same utterance while separating those from different speakers, addressing limitations of traditional non-contrastive methods. Diversity regularization alleviates the model collapse problem in non-contrastive frameworks. Comprehensive experiments demonstrate the superiority of SDPN for speaker verification.

\end{document}